\documentclass[letter]{aa}  

\usepackage{natbib}
\usepackage{lscape}
\usepackage{graphicx}
\usepackage{txfonts}
\usepackage{adjustbox}
\usepackage{threeparttable}
\usepackage{lscape}
\usepackage{color}
\usepackage{ulem}

   \usepackage[ colorlinks=true, 
                                citecolor=blue,
                                urlcolor=blue,
                                linkcolor=blue,
               pdftex,   
               hyperindex=true,  
               bookmarks=true,   
               bookmarksopen=true,   
               bookmarksnumbered=true]   
               {hyperref}   
\begin{document}

\defcitealias{2018A&A...610A..29K}{I}
\defcitealias{2019A&A...632A..28K}{II}
\defcitealias{2020A&A...642A.235K}{III}

   \title{Atmosphere of Betelgeuse before and during the Great Dimming event revealed by tomography \thanks{The reduced HERMES spectra are only available in electronic form
at the CDS via anonymous ftp to \url{cdsarc.u-strasbg.fr} (130.79.128.5) or via \url{http://cdsweb.u-strasbg.fr/cgi-bin/qcat?J/A+A/}}}

   \author{K. Kravchenko\inst{1,2}, 
          A. Jorissen\inst{3}, S. Van Eck\inst{3}, T. Merle \inst{3}, A. Chiavassa \inst{4},  
    C. Paladini\inst{1},       B. Freytag\inst{5},  B. Plez\inst{6},   \\   M. Montarg{\`e}s\inst{7,8},  \and H. Van Winckel \inst{7} 
          }

   \institute{European Southern Observatory, Alonso de Cordova 3107, Vitacura, Casilla, 19001, Santiago de Chile, Chile 
   \and 
   Max Planck Institute for extraterrestrial Physics, Giessenbachstra{\ss}e 1, D-85748 Garching, Germany \\
   \email{kkravchenko@mpe.mpg.de}
   \and
   Institut d'Astronomie et d'Astrophysique, Universit\'e Libre de Bruxelles,
              CP. 226, Boulevard du Triomphe, 1050 Bruxelles, Belgium
                          \and
             Universit\'e C\^ote d'Azur, Observatoire de la C\^ote d'Azur, CNRS, Lagrange, CS 34229, 06304 Nice Cedex 4, France 
                 \and
             Theoretical Astrophysics, Department of Physics and Astronomy at Uppsala University, Regementsv{\"a}gen 1, Box 516, SE-75120 Uppsala, Sweden   
                       \and
             Laboratoire Univers et Particules de Montpellier, Universit{\'e} de Montpellier, CNRS, 34095, Montpellier Cedex 05, France  
             \and
   Institute of Astronomy, KU Leuven, Celestijnenlaan 200D B2401, 3001 Leuven, Belgium
             \and
              LESIA, Observatoire de Paris, Universit{\'e} PSL, CNRS, Sorbonne Universit{\'e}, Universit{\'e} de Paris, 5 place Jules Janssen, 92195 Meudon, France
            }

  \abstract
{
Despite being the best studied red supergiant star in our Galaxy, the physics behind the photometric variability and mass loss of Betelgeuse is poorly understood. Moreover, recently the star has experienced an unusual fading with its visual magnitude reaching a historical minimum. The nature of this event was investigated by several studies where mechanisms, such as episodic mass loss and the presence of dark spots in the photosphere, were invoked.}
{We aim to relate the atmospheric dynamics of Betelgeuse to its photometric variability, with the main focus on the dimming event.}
{
We used the tomographic method which allowed us to probe different depths in the stellar atmosphere and to recover the corresponding disk-averaged velocity field. The method was applied to a series of high-resolution HERMES observations of Betelgeuse. Variations in the velocity field were then compared with photometric and spectroscopic variations.}
{The tomographic method reveals that 
the succession of two shocks along our line-of-sight
(in February 2018 and January 2019), the  second  one  amplifying  the  effect  of  the  first  one, combined with underlying convection and/or outward motion present at this phase of the 400~d pulsation cycle, 
produced a rapid expansion of a portion of the atmosphere of Betelgeuse and an outflow between October 2019 and February 2020. This resulted in a sudden increase in molecular opacity in the cooler upper atmosphere of Betelgeuse and, thus, in the observed unusual decrease of the star's brightness.
}
   {}

\titlerunning{Atmosphere of Betelgeuse before and during the great dimming revealed by tomography}   
\authorrunning{Kravchenko et al.}  
\maketitle

%

\section{Introduction}
\label{Sect:introduction}

Red supergiant (RSG) stars represent the late stage in the evolution of stars with initial masses larger than 8~M$_{\odot}$\footnote{The 8~M$_{\odot}$ threshold corresponds to the minimum mass required for carbon ignition in the stellar core. This value is not well established and depends on the treatment of convection in stellar evolution models \citep{2018A&ARv..26....1H}.} before they end their lives in spectacular supernova explosions. Despite the fact that RSGs were extensively studied during the last decades, important properties such as photometric variability and mass loss are still poorly constrained. Understanding these properties is crucial for a broad range of astrophysical questions including the chemical enrichment of the Galaxy, supernova progenitors, and the extragalactic distance scale \citep{2017ars..book.....L}.

Betelgeuse is one of the closest \citep[$222^{+48}_{-34}$~pc,][]{2017AJ....154...11H}, brightest (0.0 -- 1.6~mag in $V$-band\footnote{According to the American Association of Variable Star Observers (AAVSO) database \citep{AAVSO}; \url{https://aavso.org/}}), and best studied RSGs in our Galaxy. Its atmospheric parameters are listed in Table~\ref{tab:parameters}. Interferometric observations of Betelgeuse by \citet{2009A&A...508..923H}, \citet{2010A&A...515A..12C}, \citet{2011AA...529A.163O}, and \citet{2016A&A...588A.130M} have shown evidence for the presence of large convective cells in its atmosphere, which is in accordance with  predictions by \citet{1975ApJ...195..137S}. The photometric periodicity of Betelgeuse is characterized by two main periods: a short one of about 400 days and a long one of about 2000 days \citep{2006MNRAS.372.1721K}. Mechanisms such as atmospheric  pulsations in the fundamental or low-overtone modes and oscillations excited by convection cells were invoked to explain the short photometric period \citep{2006MNRAS.372.1721K}. The long photometric variations were proposed to be due to binarity or magnetic activity \citep{2004ApJ...604..800W}, or turnover of giant convection cells \citep{2010ApJ...725.1170S}.

\citet{2007A&A...469..671J} and \citet{2008AJ....135.1450G} analyzed high-resolution ($\rm R \sim 40\, 000$ and $100\, 000$, respectively) spectroscopic observations of Betelgeuse and detected time-variable Doppler shifts and line strengths. In particular, \citet{2008AJ....135.1450G} reported a hysteresis-like behavior of  the temperature (represented by a line-depth ratio) as a function of line-core velocity. It was shown that hysteresis loops turn counterclockwise, 
that is to say an increase in temperature is followed by a rise of the matter, its subsequent cooling, and falling down. This behavior was explained by the presence of large convective cells in the stellar atmosphere, which, in turn, dominate the 400-day variations of Betelgeuse. However, according to \citet[][Paper~II]{2019A&A...632A..28K}, stationary convection cannot account for the phase shifts between temperature and velocity variations observed by \citet{2008AJ....135.1450G}. Paper~\citetalias{2019A&A...632A..28K} shows that hysteresis loops are linked to the propagating shock waves, which originate from acoustic waves generated by a disturbance in the convective flow in the stellar interior.

Between October~2019 and February~2020, Betelgeuse has experienced an unusual decrease in its visual brightness by about 1~mag \citep[relative to the most recent maximum in September~2019;][see also the ESO press release\footnote{\label{note3}\url{https://www.eso.org/public/news/eso2003}; Montarg\`es et al., submitted}]{2020ATel13512....1G}, which is 
referred to as the Great Dimming event. 
Different hypotheses have been proposed to interpret the nature of this unprecedented fading. They include the formation of a dust cloud due to a recent mass-loss episode \citep[][Montarg\`es et al., submitted]{2020ApJ...891L..37L,2020ApJ...899...68D,2020RNAAS...4...39C}, thermal changes in the photosphere accompanying a temperature drop of $\sim$200~K or the  presence of a dark spot covering $\sim$50\% of the stellar surface \citep{2020ApJ...897L...9D}, or a critical transition in Betelgeuse seen as a complex dynamical system undergoing a regime shift \citep{2020A&A...640L..21G}.

The present letter aims to relate the atmospheric dynamics of Betelgeuse to its photometric variability, with the main focus on the dimming event. For this purpose, the tomographic method developed by \citet{2001A&A...379..288A} and \citet[][hereafter Paper~I]{ 2018A&A...610A..29K} was applied to time series of high-resolution spectra of Betelgeuse.

The letter is structured as follows. Section~\ref{Sect:observations} describes the spectral time series of Betelgeuse. The application to Betelgeuse of the tomographic method, which is summarized in Appendix~\ref{Sect:tomography}, is presented in Sect.~\ref{Sect:results}. The resulting interpretation of the dimming event and our conclusions are discussed in Sect.~\ref{Sect:discussion}.

\section{Observations}
\label{Sect:observations}

\begin{figure}[h!]
   \centering
   \includegraphics[width=0.5\textwidth]{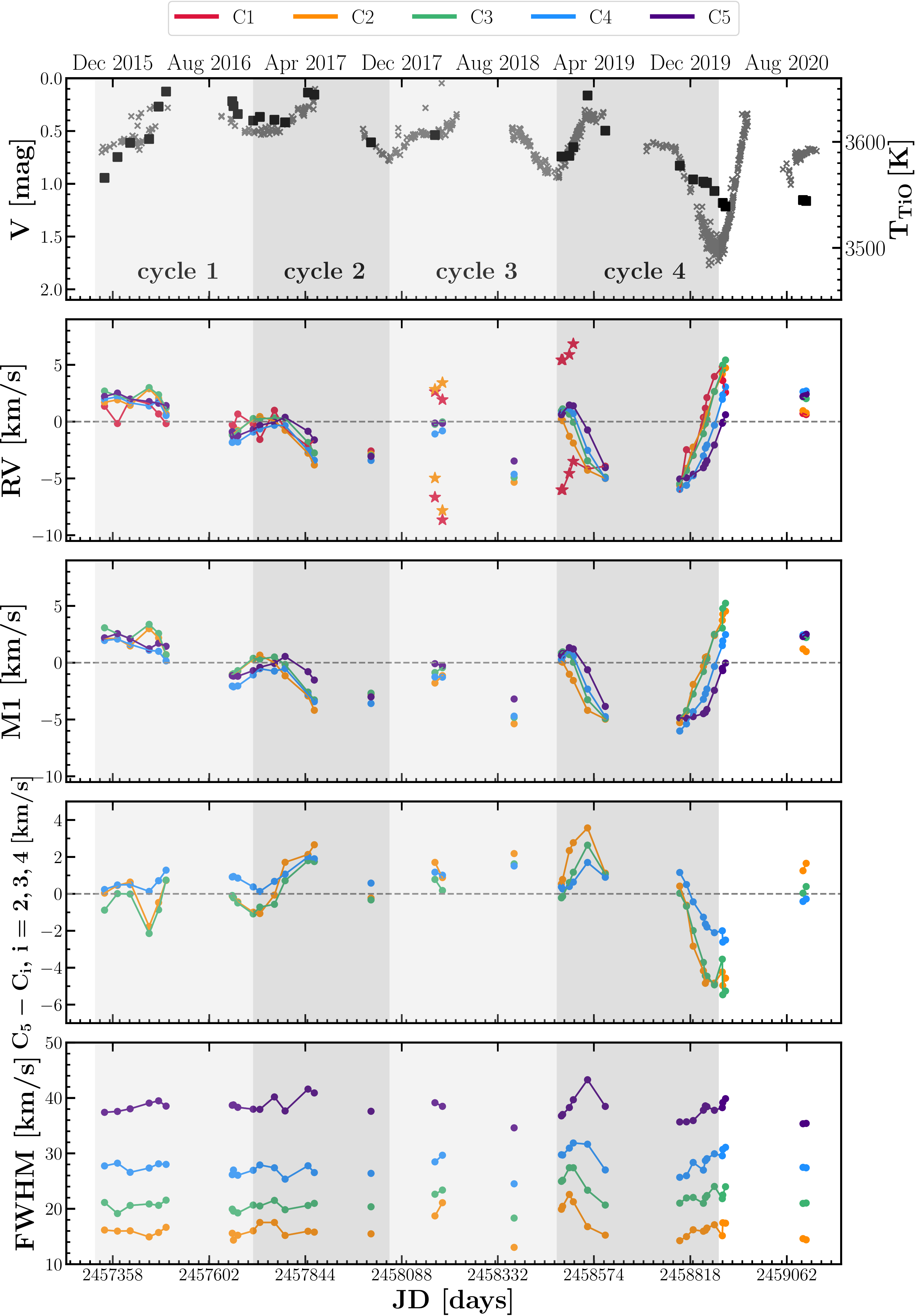}
     \caption{\textit{
     Top panel:} AAVSO visual light curve   (gray crosses). Black squares correspond to TiO-band temperatures derived from the HERMES spectra in Sect.~\ref{Sect:effective_temperatures}. Gray shaded areas in all panels define four photometric cycles, three of those being represented as hysteresis loops in Fig.~\ref{Fig:loops}. 
     \textit{Second panel from the top:}
     Radial velocities (RVs) in all masks derived from fitting Betelgeuse cross-correlation functions (CCFs) with single- or multi-component Gaussian functions. RVs from the main CCF component are connected with lines. Line-doubling events are shown with a star symbol.  
     The different colors correspond to the different masks defined in Table~\ref{tab:masks}. The same color coding is applied in the second to fifth panels. The zero point of the velocity scale corresponds to Betelgeuse center-of-mass velocity, represented by a horizontal dashed line. The uncertainty on HERMES individual RVs is on the order of 0.1~km~s$^{-1}$. \textit{Third panel:}
      $M_1$ velocities (see definition in Sect~\ref{Sect:radial_velocities}) of CCFs in masks C2--C5. 
       \textit{Fourth panel:} 
     Atmospheric velocity gradients between masks C5 and C2 (orange), C5 and C3 (green), and C5 and C4 (blue).
     \textit{Bottom panel:} The FWHM of CCFs in masks C2--C5.  
     }
         \label{Fig:RESULT}
   \end{figure}

Betelgeuse was observed with the high-resolution fiber-fed cross-dispersed echelle  spectrograph HERMES \citep{2011A&A...526A..69R} mounted on the 1.2m Mercator telescope at the Roque de Los Muchachos Observatory, La Palma (Spain). The spectral resolution of HERMES is $R= 86 000$, and the wavelength coverage is from 3800 to 9000~$\AA$. In total, 37 high-resolution, high signal-to-noise ratio (S/N~$\sim$~200-400 at 5000~$\AA$) spectra were obtained between November~2015 and September~2020, that is, covering a time span of approximately 5 years. The epochs of the HERMES observations may be located on the Betelgeuse AAVSO light curve, which can be seen in the top panel of Fig.~\ref{Fig:RESULT}. The HERMES spectra were reduced with an automated pipeline, merging the different orders  and correcting for the blaze function of the echelle grating as well as for the Earth motion around the solar-system barycenter. The long-term stability of the HERMES velocity frame is on the order of 80~m~s$^{-1}$, which was estimated from a long-term monitoring of RV standard stars \citep{2016A&A...586A.158J}.


\section{Results}
\label{Sect:results}

In this section, we apply the tomographic method of Paper~\citetalias{2018A&A...610A..29K} (summarized in  Appendix~\ref{Sect:tomography}) to the high-resolution HERMES spectra of Betelgeuse to investigate its atmospheric motions and to search for a possible link with photometric variability, with the main focus on the recent dimming event. The results are compared to those obtained by Paper~\citetalias{2019A&A...632A..28K}, where the tomographic method of Paper~\citetalias{2018A&A...610A..29K} was applied for the first time to time-series observations of another RSG star ($\mu$~Cep). 

\subsection{Cross-correlation functions (CCFs)}
\label{Sect:CCFs}

First, we cross-correlated the HERMES spectra of Betelgeuse with the set of five tomographic masks (see Table~\ref{tab:masks}) constructed in Paper~\citetalias{2019A&A...632A..28K}. Among these, the mask denoted C1 probes the innermost photospheric layer, whereas mask C5 probes the outermost layer. The resulting CCFs (displayed in Figs.~\ref{Fig:ccf_plot_1} -- \ref{Fig:ccf_plot_6}) show asymmetries in all masks. Moreover, at a few epochs (marked with a star symbol in the top panel of Fig.~\ref{Fig:RESULT}), the CCF profiles in masks C1 and C2 
are characterized by two components shifted by 10-15 $\rm km \, s^{-1}$ with respect to each other (Fig.~\ref{Fig:cycle_comparison}). To be considered meaningful, the second component must have a depth larger than $3\sigma$, where $\sigma$ is the standard deviation in the flat part of the CCFs measuring the correlation noise.  For instance, at JD~2458172 (i.e., February 22, 2018), the CCF becomes asymmetric with a faint extra component appearing on the blue wing of the main peak in masks C1 and C2. After an episode where the CCF in mask C1 shows no correlation peak (JD~2458372.7), the CCF reappears around JD~2458492 (i.e., January 8, 2019), but this time with a secondary peak on the red wing of the main peak.  The additional CCF component only appears in the innermost masks and remains weak, never equalling the contrast of the main CCF peak.

According to Paper~\citetalias{2019A&A...632A..28K}, the intensity of a CCF component is related to the size of the corresponding emitting surface on the star. For example, if a large fraction of the stellar surface is covered by rising (falling) material, then the CCF is characterized by a strong blue-shifted (red-shifted) peak. A similar scenario is observed in long-period variable (LPV) stars of Mira-type, where line doubling happens around maximum light \citep{2000A&A...362..655A}. It is associated with the appearance and passage of a spherically-symmetric shock wave across the atmosphere \citep[the Schwarzschild scenario\footnote{According to the Schwarzschild scenario, the relative strength of the blue and red components of a double line (or double CCF) varies with time and depth in the atmosphere, illustrating the upward propagation of a shock wave.},][]{Schwarzschild,2000A&A...362..655A}. In our observations of Betelgeuse, line-doubling events in February~2018 and January~2019 (also marked as star symbols in the second panel of Fig.~\ref{Fig:RESULT}) occur on the rising part of the light curve (top panel of Fig.~\ref{Fig:RESULT}). Paper~\citetalias{2019A&A...632A..28K} shows that, despite the absence of a Schwarzschild scenario in the temporal and spatial evolution of CCFs as in Mira stars, this behavior nevertheless illustrates the emergence of shock waves in the atmosphere of RSGs, but with lower amplitudes (compared to Miras) and more asymmetric shapes. The absence of similar line-doubling events at earlier epochs of our observations (i.e., during light cycles 1 and 2 in Fig.~\ref{Fig:RESULT}) means that those shock waves -- if present -- had amplitudes that were too low to be distinguishable on CCFs (see Figs.~\ref{Fig:ccf_plot_1} and \ref{Fig:ccf_plot_2}).

\subsection{Radial velocities}
\label{Sect:radial_velocities}

As a next step, we derived radial velocities (RVs) from the CCFs using the two methods described in Paper~\citetalias{2019A&A...632A..28K}. First, we fit the CCFs with a single- or multiple-component Gaussian profile using the detection of extrema (DOE) tool  \citep{2017A&A...608A..95M}. In this way, the RV information is extracted for each CCF component separately. The resulting RVs for each tomographic mask are shown in the second panel of Fig.~\ref{Fig:RESULT}. There, RVs from the single-peaked CCFs and the strongest CCF components of  double-peaked CCFs are connected with lines. The unconnected symbols correspond to the RVs of the secondary components of  double-peaked CCFs. CCFs in mask C1 are often either too shallow or too asymmetric to warrant a reliable Gaussian fit (see Figs.~\ref{Fig:ccf_plot_3} and \ref{Fig:ccf_plot_4}). Therefore, the analysis below is, in several instances, restricted to masks C2--C5. Second, RVs were computed as the first (raw) moment $M_1$ of CCFs (considering the CCF as a probability distribution, see Paper~\citetalias{2019A&A...632A..28K}). The resulting $M_1$ velocities for masks C2--C5 are shown in the third panel of Fig.~\ref{Fig:RESULT}. Both RV and $M_1$ velocities are shown in Fig.~\ref{Fig:RESULT} on a relative scale, with the zero-point set at $v_{CoM} = 20.7$~km~s$^{-1}$ corresponding to the center-of-mass (CoM) velocity of Betelgeuse ($\pm0.3$~km~s$^{-1}$), which is represented by the horizontal line in Fig.~\ref{Fig:RESULT}. This CoM velocity is the average between the values of \citet{2017ApJ...836...22H} and \citet{2018A&A...609A..67K}.

As seen in the second and third panels in Fig.~\ref{Fig:RESULT}, the maximum peak-to-peak amplitude of the RV and $M_1$ velocity variations is slightly larger in the innermost masks. A comparison with the light curve reveals that for the cycles preceding the dimming event, it amounts to $\rm 3-5 \, km \, s^{-1}$. However, during dimming, the velocity amplitude of Betelgeuse reached $\rm \sim 6 \, km \, s^{-1}$ in mask C5 and even $\rm \sim 10 \, km \, s^{-1}$ in masks C2--C3. To interpret this behavior (which cannot be ascribed to instrumental errors, which are considerably smaller, as discussed in Sect.~\ref{Sect:observations}), we defined three atmospheric velocity gradients between mask C5 and masks C2, C3, and C4 as $v_{C_5}-v_{C_i}$, where $i=2,3,4$. Velocity gradients illustrate compression ($v_{C_5}-v_{C_i} > 0$) or expansion ($v_{C_5}-v_{C_i} < 0$) of atmospheric regions over which the gradient is taken. The derived velocity gradients are displayed in the fourth panel in Fig.~\ref{Fig:RESULT} and reveal the following. First, the velocity gradients are generally nonzero, thus revealing the existence of compression or expansion within the photosphere, and they are stronger for the inner atmospheric layers. This explains the differences in the peak-to-peak amplitudes of the RV and $M_1$ velocity variations between the masks. Second, the velocity gradients very closely follow the light curve variations, with expansion corresponding to dimming
. In particular, during the dimming event, the velocity gradients are the strongest in all masks, which in turn explains the increase in the RV and $M_1$ amplitudes. Finally, during the dimming event, the velocity gradients reach values that are at least a factor of 2 larger (negative) compared to any other epoch of our observations in all masks, 
indicating rapid expansion of a portion of the atmosphere of Betelgeuse. Finally, the fact that the velocity gradients recovered their normal values in September 2020 (JD 2459100) clearly indicates that the dynamical consequences of the dimming event were then definitely over.

Figure~\ref{Fig:RESULT} also reveals that, as in $\mu$~Cep, the RV and $M_1$ velocities of Betelgeuse vary with the same periodicity as the light curve, but with a phase shift such that the light-curve maxima occur at later epochs than the velocity maxima. The phase shift amounts to about 0.2 for mask C5 and 0.3 for mask C2, and thus decreases toward the outer atmospheric layers. In particular, the difference between masks C2 and C5 is about 40 days.

\subsection{The FWHM of CCFs}

The fifth panel of Fig.~\ref{Fig:RESULT} displays the full width at half maximum (FWHM) of CCFs in different masks, which reflects the amplitude of the velocity dispersion inside a given atmospheric layer. The FWHM values increase by a factor of more than two when going from the inner to the outer atmospheric layers. The average FWHM in mask C5 is $\rm \sim 39 \, km \, s^{-1}$ while it is $\rm \sim 15 \, km \, s^{-1}$ in mask C1, whereas their uncertainty is only 0.1 - 0.2~km~s$^{-1}$. 

The FWHMs of CCFs do not show a clear correlation with the light curve variations. Nevertheless, there appears to be a phase lag between the FWHM variations in the different masks. This behavior is especially apparent between JD~2458492 (January 8, 2019) and JD~2458603 (April 29, 2019) in Fig.~\ref{Fig:RESULT}, thus somewhat before the dimming event, where a strong increase in the FWHM first occurs in mask C2 and then gradually appears in the subsequent masks. The phase lag between the masks is the same as observed in the $M_1$ velocity variations (i.e., 40 days between masks C2 and C5).

The strong increase in the FWHM visible at the beginning of Cycle~4 results from the apparition of line doubling (January 2019) in mask C1 (see the second panel of Fig.~\ref{Fig:RESULT}) at earlier epochs. The line doubling in mask C1 indicates the emergence of a shock wave at the base of the atmosphere. As the shock propagates outward, it gives rise to broader CCF profiles in the outer atmospheric layers. Thus, the phase lag of the strong increase in the FWHM during Cycle~4 along the different masks reflects the shock propagation. This interpretation is supported by the fact that the line doubling in mask C1 appears when the stellar brightness increases. In addition, the comparison with the temporal evolution of velocity $M_1$ reveals that, for a given mask, the maximum of the FWHM occurs at the exact time when the $M_1$ velocity turns negative (indicating rising material), as can be seen by comparing panels 3 and 5 of Fig.~\ref{Fig:RESULT} around January-March 2019. Finally, the identical phase lag between the $M_1$ velocity and the FWHM variations points at the same mechanism being responsible for their behavior. As pointed out in Paper~\citetalias{2019A&A...632A..28K}, physical processes linked to shocks are responsible for the $M_1$ velocity variations.

Interestingly, the above-described FWHM behavior occurring  about 6 months before the Great Dimming episode was
preceded by another episode of line doubling in February 2018 (see Sect.~\ref{Sect:CCFs}) with the FWHM of CCFs reaching about the same large values as in January 2019. The implications of these observations are discussed in Sect.~\ref{Sect:discussion}.

\subsection{TiO-band temperatures}
\label{Sect:effective_temperatures}

For each HERMES observation, we derived a temperature\footnote{According to \citet{2013ApJ...767....3D}, temperatures derived using the 1D hydrostatic model spectral fit of the TiO bands are not identical to the effective temperature of a star, which is best derived from a fit of the whole spectral energy distribution (SED). Indeed, the temperature in the outer layers (where the TiO lines form) of more realistic 3D models is lower than in their 1D counterparts, leading to increased TiO absorption. Thus, the term "temperature" used in this letter refers only to the "TiO-band" temperature.} for Betelgeuse using the method described in \citet{2017A&A...601A..10V} and Paper~\citetalias{2019A&A...632A..28K}. In short, the technique is based on the computation of the strengths of the TiO~bands having their bandheads at 5847, 6158, 6187, 7054, and 7126~$\AA$. First, the relation between the TiO-band strength and the effective temperature of a model was derived for a grid of 1D MARCS model atmospheres \citep{2008A&A...486..951G}. Then, the strengths of the TiO bands were computed for the HERMES spectra and the temperatures were deduced using the 1D-model calibration. The absolute error of our temperature determination is about 50~K, corresponding to the half-step of the effective temperature grid for MARCS models. However, the relative precision is better as may be judged from the smooth variations observed in the top panel of Fig.~\ref{Fig:RESULT}, with steps much smaller than 50~K.

The  temperatures derived in the manner described above are shown in the top panel of Fig.~\ref{Fig:RESULT} (see also Table~\ref{tab:temperatures}). A comparison with the light curve reveals that the variations of the TiO-band temperatures are generally in phase with the photometric variations. Thus, the phase shifts observed between the photometric and velocity (and the FWHM of CCF) variations apply to the  temperature variations as well. The only discrepancy between the temperature and photometric variations is observed at epochs JD~2~458~900.5 and JD~2~458~907.5, when the star recovers from the dimming event, but the temperature continues to decrease. 

We compared our  temperatures with those obtained by previous studies. \citet{2005ApJ...628..973L} and \citet{2020ApJ...891L..37L}, using moderate-resolution observations ($4-8$~\AA, corresponding to $R = 625-1250$ at $5000$~\AA), proceeded in a similar way, but by using TiO bands 
at 4761, 4954, 5167, 5448, 5847, and 6158 $\AA$ bands, thus the reddest ones were lacking. These authors found that the temperature of Betelgeuse amounts to $3650 \pm 25$~K and $3600 \pm 25$~K when $V \sim 0.5$ and $V \sim 1.6$, respectively. \citet{2020ATel13439....1G} measured the Betelgeuse surface temperature from the strength of the TiO 7190~$\AA$ bandhead and obtained 3650~K and 3565~K for $V\sim0.3$ and $V\sim1.6$, respectively. \citet{Harper2020arXiv201105982H} derived TiO-indices from the Wing narrow-band photometric filters centered at the 7190~$\AA$ TiO band and two continuum regions. Their measured TiO-temperatures amount to $3645 \pm 25$~K and $3520 \pm 25$~K in September~2019 ($V \sim 0.6$) and February~2020 ($V \sim 1.6$), respectively. According to the bottom panel of Fig.~\ref{Fig:RESULT}, for Cycle~4, at $V \sim  0.3$, 0.5, and 1.6, we find temperatures of 3645~K, 3620~K, and 3542~K, respectively. These are in good agreement (within the errorbars) with those derived by \citet{2020ATel13439....1G}, \citet{Harper2020arXiv201105982H}, and \citet{2020ApJ...891L..37L}.
The small discrepancy with the latter authors can probably be attributed to the different set of TiO bands used for the analysis, since HERMES spectra give access to the stronger red bands at 6187, 7054, and 7126~$\AA$. These bands are actually the most sensitive to temperature, as shown in Fig.~\ref{Fig:band_sensitivity}.

The comparison of our temperatures with the photometric variations shown in the top panel of Fig.~\ref{Fig:RESULT} reveals that the peak-to-peak amplitude of the temperature variations is the largest for the light cycles 1 and 4. In particular, during the rising part of Cycle 1, the temperature increases by about 81~K. Then during Cycle 4, corresponding to the dimming event, the temperature decreases by about 105~K,  which is comparable to the 125~K temperature drop found by \citealt{Harper2020arXiv201105982H}. Such a change in TiO temperature, if accompanied by a similar change in $T_{\rm eff}$, is not enough to explain the photometric variation of Betelgeuse. Indeed, the synthetic photometry of MARCS models reveals that a temperature drop of $\Delta T = 100$ K corresponds to a decrease of $\Delta V = 0.5$ mag at most, whereas the observed variation in $V$ was more than twice as large (1.3 mag).

\subsection{Hysteresis loops}

\begin{figure}
   \centering
   \includegraphics[width=9cm]{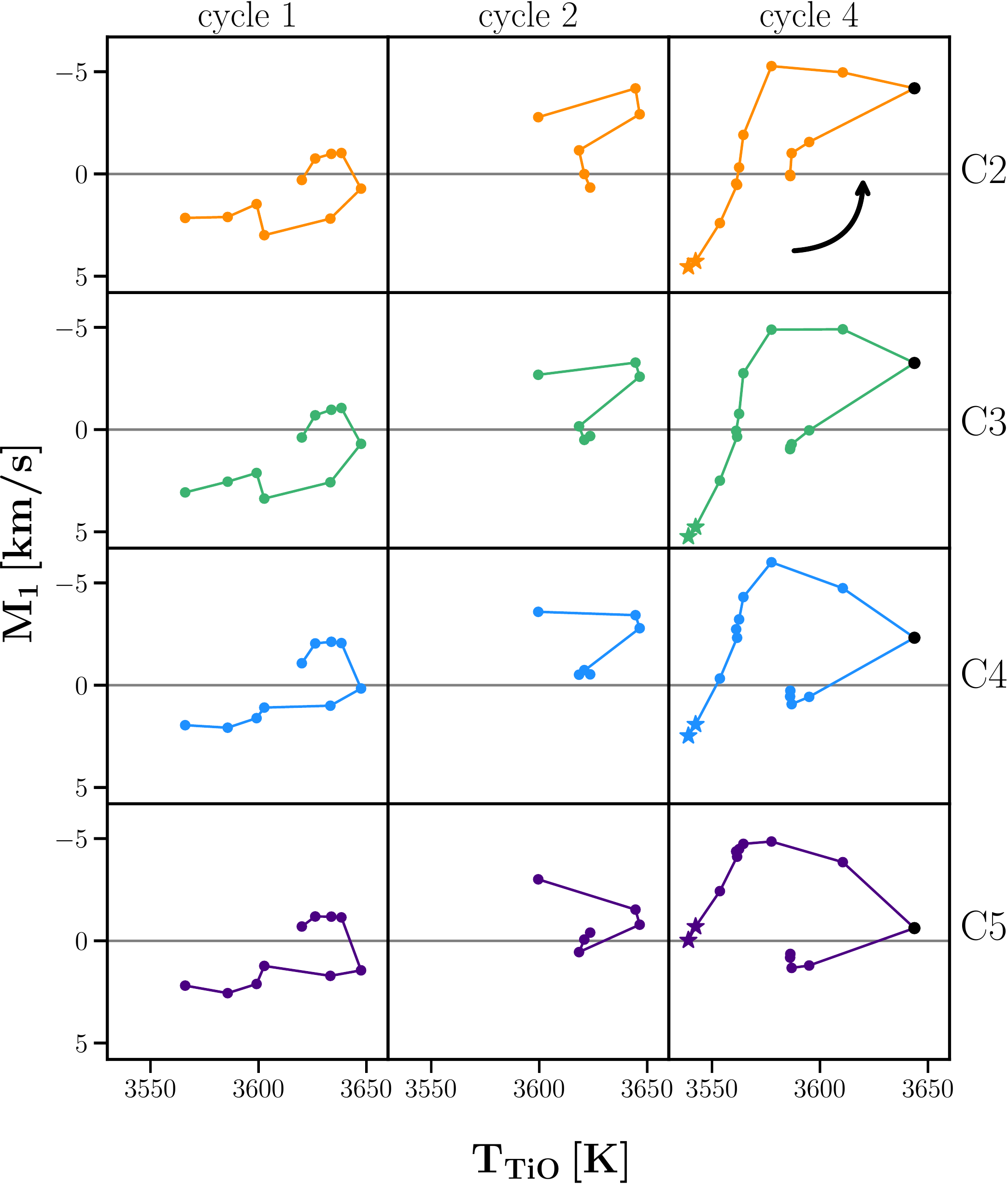}
      \caption{\textit{
      From left to right:} Hysteresis loops between TiO-band temperature and $M_1$ velocity for three pseudo light cycles of Betelgeuse from Fig.~\ref{Fig:RESULT}. \textit{From top to bottom:} Different masks, with mask C2 probing the innermost atmospheric layer. Horizontal lines in all panels indicate the center-of-mass velocity (as in Fig.~\ref{Fig:RESULT}). The arrow indicates the direction of evolution along the hysteresis loops. Star symbols in the last column correspond to epochs  JD~2~458~900.5 and JD~2~458~907.5, which are outside the shaded area defining Cycle~4 in Fig.~\ref{Fig:RESULT}. Black points in the last column denote the epoch of  maximum light preceding the dimming event.
      }
         \label{Fig:loops}
   \end{figure}

Figure~\ref{Fig:loops} displays the $M_1$ velocity in masks C2--C5 as a function of the \textcolor{blue}{TiO} temperature for the light cycles~1, 2, and 4 from Fig.~\ref{Fig:RESULT}. As in Paper~\citetalias{2019A&A...632A..28K}, the phase shift between the velocity and temperature variations translates into hysteresis loops. The star symbols along the hysteresis loop of Cycle~4 correspond to JD~2~458~900.5 and JD~2~458~907.5, that is to say two epochs following the brightness minimum of Cycle~4. The hysteresis loops in Fig.~\ref{Fig:loops} resemble those obtained earlier by \citet{2008AJ....135.1450G} and those for $\mu$~Cep from Paper~\citetalias{2019A&A...632A..28K}. According to Table~\ref{tab:betelgeuse_loops}, the characteristic timescale of hysteresis loops closely matches the short photometric period of Betelgeuse \citep[400~days,][]{2006MNRAS.372.1721K}. A similar match between the timescales of the hysteresis loops and the light variations was reported for $\mu$~Cep in Paper~\citetalias{2019A&A...632A..28K} and for Betelgeuse by \citet{2008AJ....135.1450G}. Finally, the hysteresis loops from Fig.~\ref{Fig:loops}, from \citet{2008AJ....135.1450G} and from Paper~\citetalias{2019A&A...632A..28K} show qualitative similarity in terms of their velocity and temperature ranges (see Table~\ref{tab:betelgeuse_loops} in this paper and Table~4 in Paper~\citetalias{2019A&A...632A..28K}). This further confirms  that the mechanism linked to shock-wave propagation is responsible for the velocity, temperature, and photometric variations in Betelgeuse.

\section{Discussion and conclusions}
\label{Sect:discussion}

According to the results presented in Sect.~\ref{Sect:results}, the main dynamical specificities prior to the dimming event are as follows: (i) the line-doubling episodes of February 2018 and January 2019 (second panel of Fig.~\ref{Fig:RESULT}); (ii) a concomittant increase in all CCF FWHMs (fifth panel of Fig.~\ref{Fig:RESULT}); and (iii) the relative velocity $M_1 - v_{CoM}$ turning negative during the duration of Cycle 4, which is the signature of an outflow  \citep[second and third panels of Fig.~\ref{Fig:RESULT}, as confirmed by][who also report UV line and continuum emissions in September -- November 2019 attributed to this strong outflow]{2020ApJ...899...68D}.
Assuming that the RV stayed at $-7$~km~s$^{-1}$ for 240~d in 2019 (second panel in Fig.~\ref{Fig:RESULT} and Fig.~2 of \citealt{2020ApJ...899...68D}), the distance travelled by the outflowing matter amounts to 210~R$_\odot$, which represents a small fraction of the photospheric radius of $\sim1000$~R$_\odot$ \citep[][combined with Gaia DR2]{2011AA...529A.163O}, which is certainly too small for the outflowing matter to reach the dust-formation region at $\ga 2$ stellar radii (counted from the stellar center).  
 
 Between January and March 2019, the $M_1- v_{CoM}$ velocity turned negative for each mask at the exact time the FWHM maximum reached the corresponding layer. Moreover,  the upward propagation of the broad CCFs during Cycle 4 
 (fifth panel of Fig.~\ref{Fig:RESULT}) reveals the  propagation of the shock wave through the atmosphere. Thus, it is clear that a strong shock has appeared at the bottom of the photosphere in January 2019 and it has moved upward, which first led to a strong compression ($v_{C_5} - v_{C_i} >0$) and then, as the dimming started, to a strong expansion ($v_{C_5} - v_{C_i}< 0$) of the atmosphere. Observational evidence is strong to link this  compression (expansion) in the photosphere with the increase (decrease) in the TiO temperature. In particular, during the strong compression episode (seen in January -- May 2019 in panel~4 of Fig.~\ref{Fig:RESULT}), the temperature shows a clear maximum as expected when a perfect gas is being compressed \citep[e.g., Fig.~6 of][]{2016A&A...589A.130L}. Conversely, during the expansion episode, the temperature decreases as expected. The emergence of a similar  shock in February 2018, with even larger upward velocities than in January 2019,  is likely; although, it is unfortunately not documented by observations as well.

Thus the extraordinary strong upflow occurring during the duration of
Cycle~4 results from the conjunction of several effects: (i) 
the succession of two shocks along our line-of-sight, the second one amplifying the effect of the first one; (ii) a specific convective pattern with mostly rising matter on the visible hemisphere of Betelgeuse, with a  large  cell\footnote{Ultraviolet observations of Betelgeuse by \citet{2020ApJ...899...68D} confirm that  a bright and hot structure (rising matter in a large convective cell) has appeared in the southern hemisphere of Betelgeuse between September and November 2019.} resulting from random fluctuations in the stellar interior; and (iii)
the previous phenomena amplified by the coincidence with outward motion present at this phase of the $\sim 400$~d pulsation cycle. We interpret the Great Dimming event as a consequence of this strong outflow which has caused a cooling of the outer layers,  resulting in a sudden increase in molecular opacity. This phenomenon has been described in the literature as "molecular plumes" rising from the photosphere of supergiants \citep[][Freytag et al., in prep.]{Kervella2018A&A...609A..67K}, or "molecular reservoirs" \citep{Harper2020arXiv201105982H}, a nonspherical version of the "molsphere" observed around AGB stars \citep{Freytag2017A&A...600A.137F}. 

This scenario is supported by the VLT/SPHERE measurements (see footnote~\ref{note3}), which revealed an obscuration in the southern half of Betelgeuse disk in the optical in January 2020. Whether or not this obscuration was caused by dust is a crucial point. Contrary to expectations, when new dust is produced, no significant flux increase has been recorded in the IR \citep[][and references therein]{Harper2020arXiv201105982H}.
Moreover this obscuration event has been recorded
(20\% dimming) in the submillimeter range \citep{2020ApJ...897L...9D}; however, new dust is not expected to significantly change the spectrum at wavelengths larger than $100~\mu$m. Therefore, similarly to \citet{Harper2020arXiv201105982H}, we conclude that dust does not appear to be a crucial ingredient  to account for the Great Dimming.

\begin{acknowledgements}

This work is based on observations obtained with the HERMES spectrograph, which is supported by the Fund for Scientific Research of Flanders (FWO), Belgium, the Research Council of K.U.~Leuven, Belgium, the Fonds de la Recherche Scientifique (F.R.S.-FNRS), Belgium, the Royal Observatory of Belgium, the Observatoire de Gen{\`e}ve, Switzerland and the Th{\"u}ringer Landessternwarte Tautenburg, Germany. We acknowledge with thanks the variable star observations from the AAVSO International Database contributed by observers worldwide and used in this research. S.V.E. and T.M. thank {\it Fondation ULB} for its support. MM acknowledge support from the ERC consolidator grant 646758 AEROSOL. BF acknowledges support from the Swedish Research Council (Vetenskapsr{\aa}det) and the ERC Advanced Grant 883867 EXWINGS.

\end{acknowledgements}

\bibliographystyle{aa}
\bibliography{bibliography}

\begin{thebibliography}{43}
\expandafter\ifx\csname natexlab\endcsname\relax\def\natexlab#1{#1}\fi

\bibitem[{{Alvarez} {et~al.}(2000){Alvarez}, {Jorissen}, {Plez}, {Gillet}, \&
  {Fokin}}]{2000A&A...362..655A}
{Alvarez}, R., {Jorissen}, A., {Plez}, B., {Gillet}, D., \& {Fokin}, A. 2000,
  \aap, 362, 655

\bibitem[{{Alvarez} {et~al.}(2001){Alvarez}, {Jorissen}, {Plez}, {Gillet},
  {Fokin}, \& {Dedecker}}]{2001A&A...379..288A}
{Alvarez}, R., {Jorissen}, A., {Plez}, B., {et~al.} 2001, \aap, 379, 288

\bibitem[{{Chiavassa} {et~al.}(2010){Chiavassa}, {Haubois}, {Young}, {Plez},
  {Josselin}, {Perrin}, \& {Freytag}}]{2010A&A...515A..12C}
{Chiavassa}, A., {Haubois}, X., {Young}, J.~S., {et~al.} 2010, \aap, 515, A12

\bibitem[{{Cotton} {et~al.}(2020){Cotton}, {Bailey}, {Horta}, {Norris}, \&
  {Lomax}}]{2020RNAAS...4...39C}
{Cotton}, D.~V., {Bailey}, J., {Horta}, A.~D., {Norris}, B. R.~M., \& {Lomax},
  J.~R. 2020, Research Notes of the American Astronomical Society, 4, 39

\bibitem[{{Davies} {et~al.}(2013){Davies}, {Kudritzki}, {Plez}, {Trager},
  {Lan{\c{c}}on}, {Gazak}, {Bergemann}, {Evans}, \&
  {Chiavassa}}]{2013ApJ...767....3D}
{Davies}, B., {Kudritzki}, R.-P., {Plez}, B., {et~al.} 2013, \apj, 767, 3

\bibitem[{{Dharmawardena} {et~al.}(2020){Dharmawardena}, {Mairs}, {Scicluna},
  {Bell}, {McDonald}, {Menten}, {Weiss}, \& {Zijlstra}}]{2020ApJ...897L...9D}
{Dharmawardena}, T.~E., {Mairs}, S., {Scicluna}, P., {et~al.} 2020, \apjl, 897,
  L9

\bibitem[{{Dolan} {et~al.}(2016){Dolan}, {Mathews}, {Lam}, {Quynh Lan},
  {Herczeg}, \& {Dearborn}}]{2016ApJ...819....7D}
{Dolan}, M.~M., {Mathews}, G.~J., {Lam}, D.~D., {et~al.} 2016, \apj, 819, 7

\bibitem[{{Dupree} {et~al.}(2020){Dupree}, {Strassmeier}, {Matthews},
  {Uitenbroek}, {Calderwood}, {Granzer}, {Guinan}, {Leike}, {Montarg{\`e}s},
  {Richards}, {Wasatonic}, \& {Weber}}]{2020ApJ...899...68D}
{Dupree}, A.~K., {Strassmeier}, K.~G., {Matthews}, L.~D., {et~al.} 2020, \apj,
  899, 68

\bibitem[{{Freytag} {et~al.}(2017){Freytag}, {Liljegren}, \&
  {H{\"o}fner}}]{Freytag2017A&A...600A.137F}
{Freytag}, B., {Liljegren}, S., \& {H{\"o}fner}, S. 2017, \aap, 600, A137

\bibitem[{{George} {et~al.}(2020){George}, {Kachhara}, {Misra}, \&
  {Ambika}}]{2020A&A...640L..21G}
{George}, S.~V., {Kachhara}, S., {Misra}, R., \& {Ambika}, G. 2020, \aap, 640,
  L21

\bibitem[{{Gray}(2008)}]{2008AJ....135.1450G}
{Gray}, D.~F. 2008, \aj, 135, 1450

\bibitem[{{Guinan} {et~al.}(2020){Guinan}, {Wasatonic}, {Calderwood}, \&
  {Carona}}]{2020ATel13512....1G}
{Guinan}, E., {Wasatonic}, R., {Calderwood}, T., \& {Carona}, D. 2020, The
  Astronomer's Telegram, 13512, 1

\bibitem[{{Guinan} \& {Wasatonic}(2020)}]{2020ATel13439....1G}
{Guinan}, E.~F. \& {Wasatonic}, R.~J. 2020, The Astronomer's Telegram, 13439, 1

\bibitem[{{Gustafsson} {et~al.}(2008){Gustafsson}, {Edvardsson}, {Eriksson},
  {J{\o}rgensen}, {Nordlund}, \& {Plez}}]{2008A&A...486..951G}
{Gustafsson}, B., {Edvardsson}, B., {Eriksson}, K., {et~al.} 2008, \aap, 486,
  951

\bibitem[{{Harper} {et~al.}(2008){Harper}, {Brown}, \&
  {Guinan}}]{2008AJ....135.1430H}
{Harper}, G.~M., {Brown}, A., \& {Guinan}, E.~F. 2008, \aj, 135, 1430

\bibitem[{{Harper} {et~al.}(2017{\natexlab{a}}){Harper}, {Brown}, {Guinan},
  {O'Gorman}, {Richards}, {Kervella}, \& {Decin}}]{2017AJ....154...11H}
{Harper}, G.~M., {Brown}, A., {Guinan}, E.~F., {et~al.} 2017{\natexlab{a}},
  \aj, 154, 11

\bibitem[{{Harper} {et~al.}(2017{\natexlab{b}}){Harper}, {DeWitt}, {Richter},
  {Greathouse}, {Ryde}, {Guinan}, {O'Gorman}, \& {Vacca}}]{2017ApJ...836...22H}
{Harper}, G.~M., {DeWitt}, C., {Richter}, M.~J., {et~al.} 2017{\natexlab{b}},
  \apj, 836, 22

\bibitem[{{Harper} {et~al.}(2020){Harper}, {Guinan}, {Wasatonic}, \&
  {Ryde}}]{Harper2020arXiv201105982H}
{Harper}, G.~M., {Guinan}, E.~F., {Wasatonic}, R., \& {Ryde}, N. 2020, arXiv
  e-prints, arXiv:2011.05982

\bibitem[{{Haubois} {et~al.}(2009){Haubois}, {Perrin}, {Lacour}, {Verhoelst},
  {Meimon}, {Mugnier}, {Thi{\'e}baut}, {Berger}, {Ridgway}, {Monnier},
  {Millan-Gabet}, \& {Traub}}]{2009A&A...508..923H}
{Haubois}, X., {Perrin}, G., {Lacour}, S., {et~al.} 2009, \aap, 508, 923

\bibitem[{{H{\"o}fner} \& {Olofsson}(2018)}]{2018A&ARv..26....1H}
{H{\"o}fner}, S. \& {Olofsson}, H. 2018, \aapr, 26, 1

\bibitem[{{Jorissen} {et~al.}(2016){Jorissen}, {Van Eck}, {Van Winckel},
  {Merle}, {Boffin}, {Andersen}, {Nordstr{\"o}m}, {Udry}, {Masseron},
  {Lenaerts}, \& {Waelkens}}]{2016A&A...586A.158J}
{Jorissen}, A., {Van Eck}, S., {Van Winckel}, H., {et~al.} 2016, \aap, 586,
  A158

\bibitem[{{Josselin} \& {Plez}(2007)}]{2007A&A...469..671J}
{Josselin}, E. \& {Plez}, B. 2007, \aap, 469, 671

\bibitem[{{Kafka}(2018)}]{AAVSO}
{Kafka}, S. 2018, Observations from the AAVSO International Database,
  https://www.aavso.org

\bibitem[{{Kervella} {et~al.}(2018{\natexlab{a}}){Kervella}, {Decin},
  {Richards}, {Harper}, {McDonald}, {O'Gorman}, {Montarg{\`e}s}, {Homan}, \&
  {Ohnaka}}]{2018A&A...609A..67K}
{Kervella}, P., {Decin}, L., {Richards}, A. M.~S., {et~al.} 2018{\natexlab{a}},
  \aap, 609, A67

\bibitem[{{Kervella} {et~al.}(2018{\natexlab{b}}){Kervella}, {Decin},
  {Richards}, {Harper}, {McDonald}, {O'Gorman}, {Montarg{\`e}s}, {Homan}, \&
  {Ohnaka}}]{Kervella2018A&A...609A..67K}
{Kervella}, P., {Decin}, L., {Richards}, A.~M.~S., {et~al.} 2018{\natexlab{b}},
  \aap, 609, A67

\bibitem[{{Kiss} {et~al.}(2006){Kiss}, {Szab{\'o}}, \&
  {Bedding}}]{2006MNRAS.372.1721K}
{Kiss}, L.~L., {Szab{\'o}}, G.~M., \& {Bedding}, T.~R. 2006, \mnras, 372, 1721

\bibitem[{{Kravchenko} {et~al.}(2019){Kravchenko}, {Chiavassa}, {Van Eck},
  {Jorissen}, {Merle}, {Freytag}, \& {Plez}}]{2019A&A...632A..28K}
{Kravchenko}, K., {Chiavassa}, A., {Van Eck}, S., {et~al.} 2019, \aap, 632, A28

\bibitem[{{Kravchenko} {et~al.}(2018){Kravchenko}, {Van Eck}, {Chiavassa},
  {Jorissen}, {Freytag}, \& {Plez}}]{2018A&A...610A..29K}
{Kravchenko}, K., {Van Eck}, S., {Chiavassa}, A., {et~al.} 2018, \aap, 610, A29

\bibitem[{{Kravchenko} {et~al.}(2020){Kravchenko}, {Wittkowski}, {Jorissen},
  {Chiavassa}, {Van Eck}, {Anderson}, {Freytag}, \&
  {K{\"a}ufl}}]{2020A&A...642A.235K}
{Kravchenko}, K., {Wittkowski}, M., {Jorissen}, A., {et~al.} 2020, \aap, 642,
  A235

\bibitem[{{Le Bertre} {et~al.}(2012){Le Bertre}, {Matthews}, {G{\'e}rard}, \&
  {Libert}}]{2012MNRAS.422.3433L}
{Le Bertre}, T., {Matthews}, L.~D., {G{\'e}rard}, E., \& {Libert}, Y. 2012,
  \mnras, 422, 3433

\bibitem[{{Levesque}(2017)}]{2017ars..book.....L}
{Levesque}, E.~M. 2017, {Astrophysics of Red Supergiants} ({IOP Publishing})

\bibitem[{{Levesque} \& {Massey}(2020)}]{2020ApJ...891L..37L}
{Levesque}, E.~M. \& {Massey}, P. 2020, \apjl, 891, L37

\bibitem[{{Levesque} {et~al.}(2005){Levesque}, {Massey}, {Olsen}, {Plez},
  {Josselin}, {Maeder}, \& {Meynet}}]{2005ApJ...628..973L}
{Levesque}, E.~M., {Massey}, P., {Olsen}, K.~A.~G., {et~al.} 2005, \apj, 628,
  973

\bibitem[{{Liljegren} {et~al.}(2016){Liljegren}, {H{\"o}fner}, {Nowotny}, \&
  {Eriksson}}]{2016A&A...589A.130L}
{Liljegren}, S., {H{\"o}fner}, S., {Nowotny}, W., \& {Eriksson}, K. 2016, \aap,
  589, A130

\bibitem[{{Merle} {et~al.}(2017){Merle}, {Van Eck}, {Jorissen}, {Van der
  Swaelmen}, {Masseron}, {Zwitter}, {Hatzidimitriou}, {Klutsch}, {Pourbaix},
  {Blomme}, {Worley}, {Sacco}, {Lewis}, {Abia}, {Traven}, {Sordo}, {Bragaglia},
  {Smiljanic}, {Pancino}, {Damiani}, {Hourihane}, {Gilmore}, {Randich},
  {Koposov}, {Casey}, {Morbidelli}, {Franciosini}, {Magrini}, {Jofre},
  {Costado}, {Jeffries}, {Bergemann}, {Lanzafame}, {Bayo}, {Carraro},
  {Flaccomio}, {Monaco}, \& {Zaggia}}]{2017A&A...608A..95M}
{Merle}, T., {Van Eck}, S., {Jorissen}, A., {et~al.} 2017, \aap, 608, A95

\bibitem[{{Montarg{\`e}s} {et~al.}(2016){Montarg{\`e}s}, {Kervella}, {Perrin},
  {Chiavassa}, {Le Bouquin}, {Auri{\`e}re}, {L{\'o}pez Ariste}, {Mathias},
  {Ridgway}, {Lacour}, {Haubois}, \& {Berger}}]{2016A&A...588A.130M}
{Montarg{\`e}s}, M., {Kervella}, P., {Perrin}, G., {et~al.} 2016, \aap, 588,
  A130

\bibitem[{{Ohnaka} {et~al.}(2011){Ohnaka}, {Weigelt}, {Millour}, {Hofmann},
  {Driebe}, {Schertl}, {Chelli}, {Massi}, {Petrov}, \&
  {Stee}}]{2011AA...529A.163O}
{Ohnaka}, K., {Weigelt}, G., {Millour}, F., {et~al.} 2011, \aap, 529, A163

\bibitem[{{Raskin} {et~al.}(2011){Raskin}, {van Winckel}, {Hensberge},
  {Jorissen}, {Lehmann}, {Waelkens}, {Avila}, {de Cuyper}, {Degroote},
  {Dubosson}, {Dumortier}, {Fr{\'e}mat}, {Laux}, {Michaud}, {Morren}, {Perez
  Padilla}, {Pessemier}, {Prins}, {Smolders}, {van Eck}, \&
  {Winkler}}]{2011A&A...526A..69R}
{Raskin}, G., {van Winckel}, H., {Hensberge}, H., {et~al.} 2011, \aap, 526, A69

\bibitem[{Schwarzschild(1954)}]{Schwarzschild}
Schwarzschild, M. 1954, in Transactions of the International Astronomical
  Union, Vol.~8 (Cambridge University Press), 811–812

\bibitem[{{Schwarzschild}(1975)}]{1975ApJ...195..137S}
{Schwarzschild}, M. 1975, \apj, 195, 137

\bibitem[{{Stothers}(2010)}]{2010ApJ...725.1170S}
{Stothers}, R.~B. 2010, \apj, 725, 1170

\bibitem[{{Van Eck} {et~al.}(2017){Van Eck}, {Neyskens}, {Jorissen}, {Plez},
  {Edvardsson}, {Eriksson}, {Gustafsson}, {J{\o}rgensen}, \&
  {Nordlund}}]{2017A&A...601A..10V}
{Van Eck}, S., {Neyskens}, P., {Jorissen}, A., {et~al.} 2017, \aap, 601, A10

\bibitem[{{Wood} {et~al.}(2004){Wood}, {Olivier}, \&
  {Kawaler}}]{2004ApJ...604..800W}
{Wood}, P.~R., {Olivier}, E.~A., \& {Kawaler}, S.~D. 2004, \apj, 604, 800

\end{thebibliography}

\begin{appendix}

\section{The tomographic method}
\label{Sect:tomography}

The tomographic method allows one to probe different geometrical depths in the stellar atmosphere and to recover the surface-averaged line-of-sight velocity at those various geometrical depths \citep[][and Paper~\citetalias{2018A&A...610A..29K}]{2001A&A...379..288A}. The technique is based on sorting different spectral lines according to their formation depth, which is provided by the maximum of the line-depth contribution function \citepalias[CF; see Fig.~1 of  Paper][]{2018A&A...610A..29K}. The formation depth of spectral lines is expressed in an optical-depth scale computed at a reference wavelength (e.g., $\lambda$~=~5000~$\AA$) built from continuum opacities only according to Eq.~3 of Paper~\citetalias{2018A&A...610A..29K}. This reference optical depth serves as a proxy for geometrical depth. A set of spectral templates (so-called masks) was constructed so as to identify spectral lines forming inside a given range of reference optical depths (or equivalently, geometrical depths) in the atmosphere, as listed in Table~\ref{tab:masks}. Cross-correlating these masks with a stellar spectrum provides information on the velocity field, the average shape, and strength of lines for a series of atmospheric slices. The capability of the tomographic method to correctly recover the line-of-sight velocity at different depths in a stellar atmosphere was validated in Paper~\citetalias{2018A&A...610A..29K} using state-of-the-art 3D~RHD simulations of convection in an RSG  atmosphere.

The tomographic method  was first applied to the spectroscopic observations of Mira stars in order to interpret their atmospheric dynamics  \citep{2001A&A...379..288A}. It was shown that, around maximum light, the line profiles of Mira stars appear double-peaked and follow the Schwarzschild scenario, reflecting the upward propagation of a spherically-symmetric shock front across the photosphere. The tomographic method was then applied by \citet{2007A&A...469..671J} to time-series spectroscopic observations of RSG stars, which revealed complex time-variable upward and downward motions in their atmospheres.
Paper~\citetalias{2019A&A...632A..28K} applied an improved version of the tomographic method to the RSG star $\mu$~Cep. This study revealed that the spatial and temporal evolution of $\mu$~Cep CCFs does not follow the Schwarzschild scenario as in Mira variables due to smaller-amplitude velocity fields, the absence of spherically-symmetric shock waves, and more extended atmospheres.

Finally, \citet{2020A&A...642A.235K} applied the tomographic technique to the high-resolution spectro-interferometric VLTI/AMBER observations of the Mira-type star S~Ori and validated the capability of tomography to probe different geometrical depths in the stellar atmosphere. Moreover, this study derived, for the first time, a quantitative relation between optical- and geometrical-depth scales for S Ori.

The various applications of the tomographic method reported above use different sets of masks, constructed from 1D static model atmospheres matching the atmospheric parameters of the stars under consideration. 
The set of masks used here (Table~\ref{tab:masks}) is the same as that built in Paper~\citetalias{2019A&A...632A..28K} for RSGs, based on a 1D MARCS model with $T_{\rm eff} = 3400$~K and $\log g = -0.4$, which falls in the parameter range of RSGs \citep{2005ApJ...628..973L}. The model was extended to $\log \tau_0 = -4.6$ with respect to the standard version of MARCS models (limited to $\log \tau_0 = -3.5$) in order to properly handle the lines forming in the outermost layers (mask C5). For the sake of testing the sensitivity of the masks 
to the adopted model, another set of masks was built from a  model with $T_{\rm eff} = 3600$~K and $\log g =  0$. The vast majority of lines ($> 88$\% per mask) are assigned to the same mask in both models. Therefore, the small   mismatch between the atmospheric  parameters of Betelgeuse (Table~\ref{tab:parameters}) and those of the  model that were used to build the set of masks from Paper~\citetalias{2019A&A...632A..28K}  (Table~\ref{tab:masks}) does not jeopardize the conclusions reached in the present work.

\begin{table}
\begin{center}
\begin{threeparttable}
\caption[]{Properties of the tomographic masks constructed in Paper~\citetalias{2019A&A...632A..28K} from a 1D MARCS model with $T_{\rm eff} = 3400$~K and $\log g = -0.4$.}
\label{tab:masks} 
\begin{tabular}{c c c }
\hline \hline
\noalign{\smallskip}
Mask  & $\rm \log \tau_0$ limits\tnote{*} & number of lines   \\
\noalign{\smallskip}
\hline
\noalign{\smallskip}
C1 & $ -1.0 < \log \tau_0 \leq 0.0 $  & 419 \\
C2 & $ -2.0 < \log \tau_0 \leq -1.0 $ & 1750 \\
C3 & $ -3.0 < \log \tau_0 \leq -2.0 $ & 1199 \\
C4 & $ -4.0 < \log \tau_0 \leq -3.0 $ & 433 \\
C5 & $ -4.6 < \log \tau_0 \leq -4.0 $           & 378 \\
\noalign{\smallskip}
\hline
\end{tabular}
\begin{tablenotes}
\item [*] $\tau_0$ is the reference optical depth computed with the continuum opacity at $\lambda~=~5000$~\AA\  and it serves as a proxy for geometrical depth. 
\end{tablenotes}
\end{threeparttable}
\end{center}
\end{table}

\section{Additional material}

\begin{table}
\begin{center}
\caption[]{Fundamental stellar parameters of Betelgeuse.}
\label{tab:parameters} 
\begin{tabular}{c c c }
\hline \hline
\noalign{\smallskip}
Parameter  &  Value & Reference   \\
\noalign{\smallskip}
\hline
\noalign{\smallskip}
$\rm T_{\rm eff} \, [K]$           & $3650 \pm 25$          & \citet{2005ApJ...628..973L}  \\
$\log g$ [c.g.s.]              &   0.08                &  {\color{blue} Josselin \& Plez (2007)}  \\
Initial mass $\rm [M_{\odot}]$      & 20             & \citet{2016ApJ...819....7D} \\
$\rm Radius \, [R_{\odot}]$    &  1021               &  \citet{2011AA...529A.163O}   \\
$\log (L$/$\rm L_{\odot})$         & $5.10 \pm 0.22$        & \citet{2008AJ....135.1430H} \\
Mass loss [$\rm M_{\odot} \, \rm yr^{-1}$]    & $1.2 \times 10^{-6}$   &  \citet{2012MNRAS.422.3433L} \\
\noalign{\smallskip}
\hline
\end{tabular}
\end{center}
\end{table}

\begin{table*}[h!]
\begin{center}
\begin{threeparttable}
\caption[]{Properties of hysteresis loops of Betelgeuse from the present paper and \citet{2008AJ....135.1450G}.}
\label{tab:betelgeuse_loops} 
\begin{tabular}{c c c c c c}
\hline \hline
\noalign{\smallskip}
  & mask & RV range  & {$(RV_{\rm min}+RV_{\rm max})/2$} & $T$ range  & Timescale\\
  & & (km/s)& (km/s) & (K) & (d)  \\
  
\noalign{\smallskip}
\hline
\noalign{\smallskip}
         & C2 & 4.0 & 1.0   &     &      \\
cycle 1  & C3 & 4.4 & 1.2   & 81  &  400 \\
         & C4 & 4.2 & 0.0   &     &      \\
         & C5 & 3.7 & 0.7   &     &      \\
\hline \\        
         & C2 & 4.9 & -1.8  &     &      \\       
cycle 2  & C3 & 3.8 & -1.4  & 45  &  345 \\
         & C4 & 3.1 & -2.1  &     &      \\
         & C5 & 3.6 & -1.2  &     &      \\
\hline \\       
         & C2 & 7.7 (9.8)\tnote{a}  & -1.4 (-0.4)  &          &      \\
cycle 4  & C3 & 7.4 (10.1) & -1.2 (0.16)  & 91 (105) &  410 \\
         & C4 & 6.9 (8.5)  & -2.5 (-1.8)  &          &      \\
         & C5 & 6.2 (6.2)  & -1.8 (-1.8)  &          &      \\ 
\hline\\
\citet{2008AJ....135.1450G} &  & 4-8 & & up to $\sim$100 & $\sim$ 400 \\    
\noalign{\smallskip}
\hline
\end{tabular}
\begin{tablenotes}
\item [a] Values in parentheses correspond to measurements from hysteresis loops of Cycle~4 with two additional epochs JD~2~458~900.5 and JD~2~458~907.5 shown as star symbols in Fig.~\ref{Fig:loops}.
\end{tablenotes}
\end{threeparttable}
\end{center}
\end{table*}

   \begin{figure}
   \centering
   \includegraphics[width=8cm]{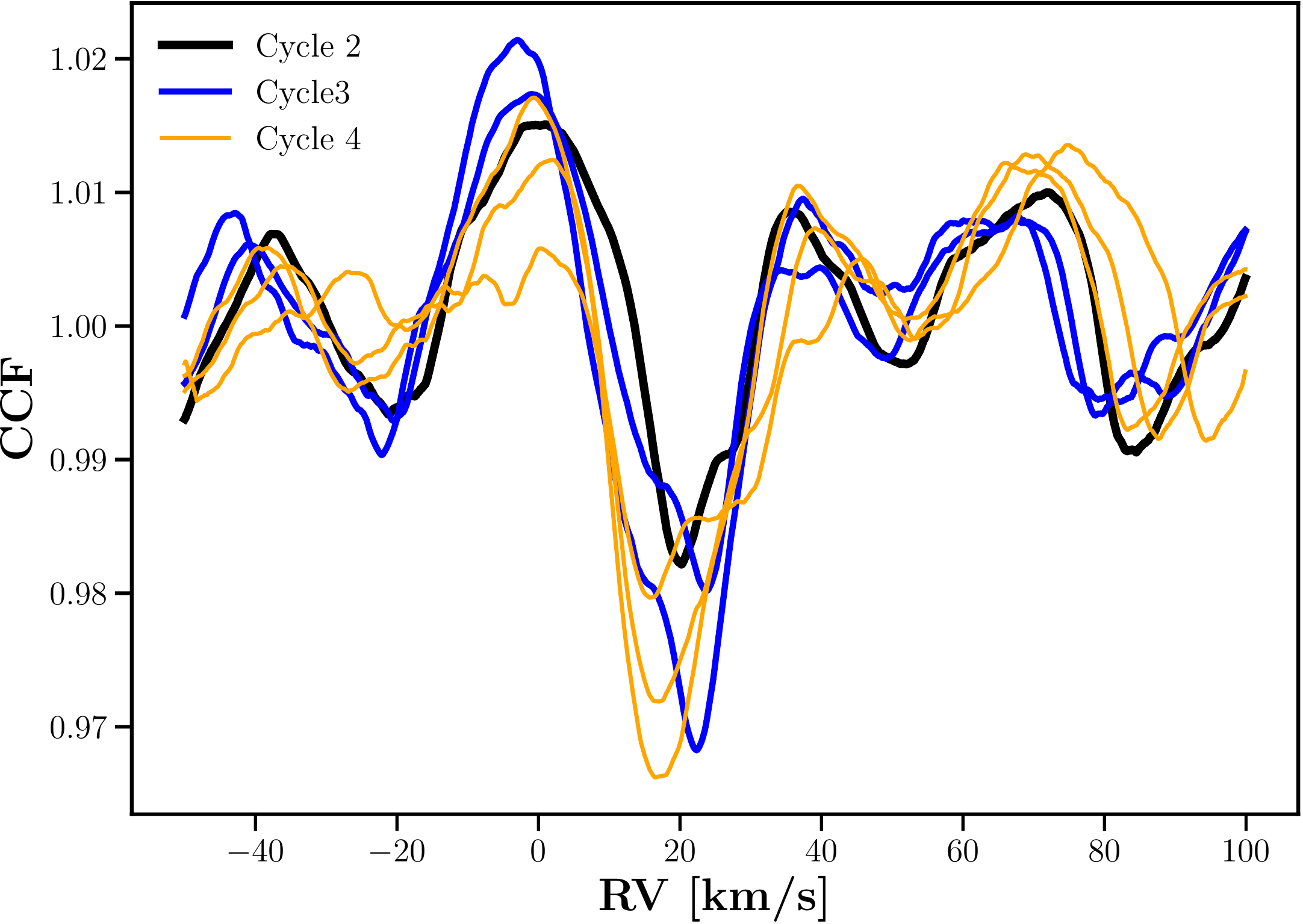}\\
   \includegraphics[width=8cm]{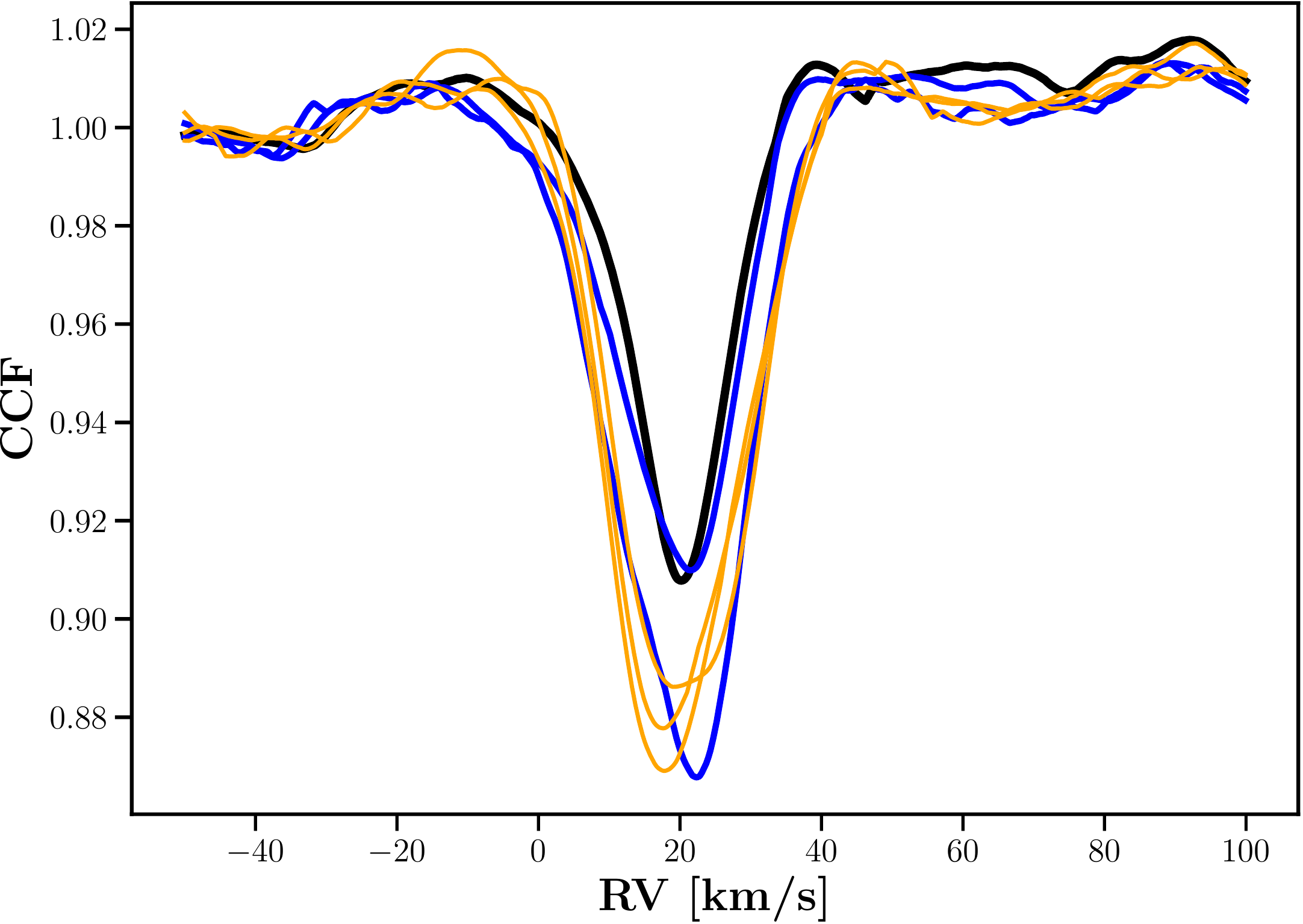} 
      \caption{\textit{Top panel:} CCFs in mask C1 corresponding to the line-doubling phases of Cycles 2 (black line; 2017-02-08), 3 (blue lines; 2018-02-22, 2018-03-13), and 4 (orange lines; 2019-01-11, 2019-01-28, 2019-02-08) from Fig.~\ref{Fig:RESULT}. \textit{Bottom panel:} Same as top panel, but for mask C2.
      }
         \label{Fig:cycle_comparison}
   \end{figure}

   \begin{figure}
   \centering
   \includegraphics[width=9cm]{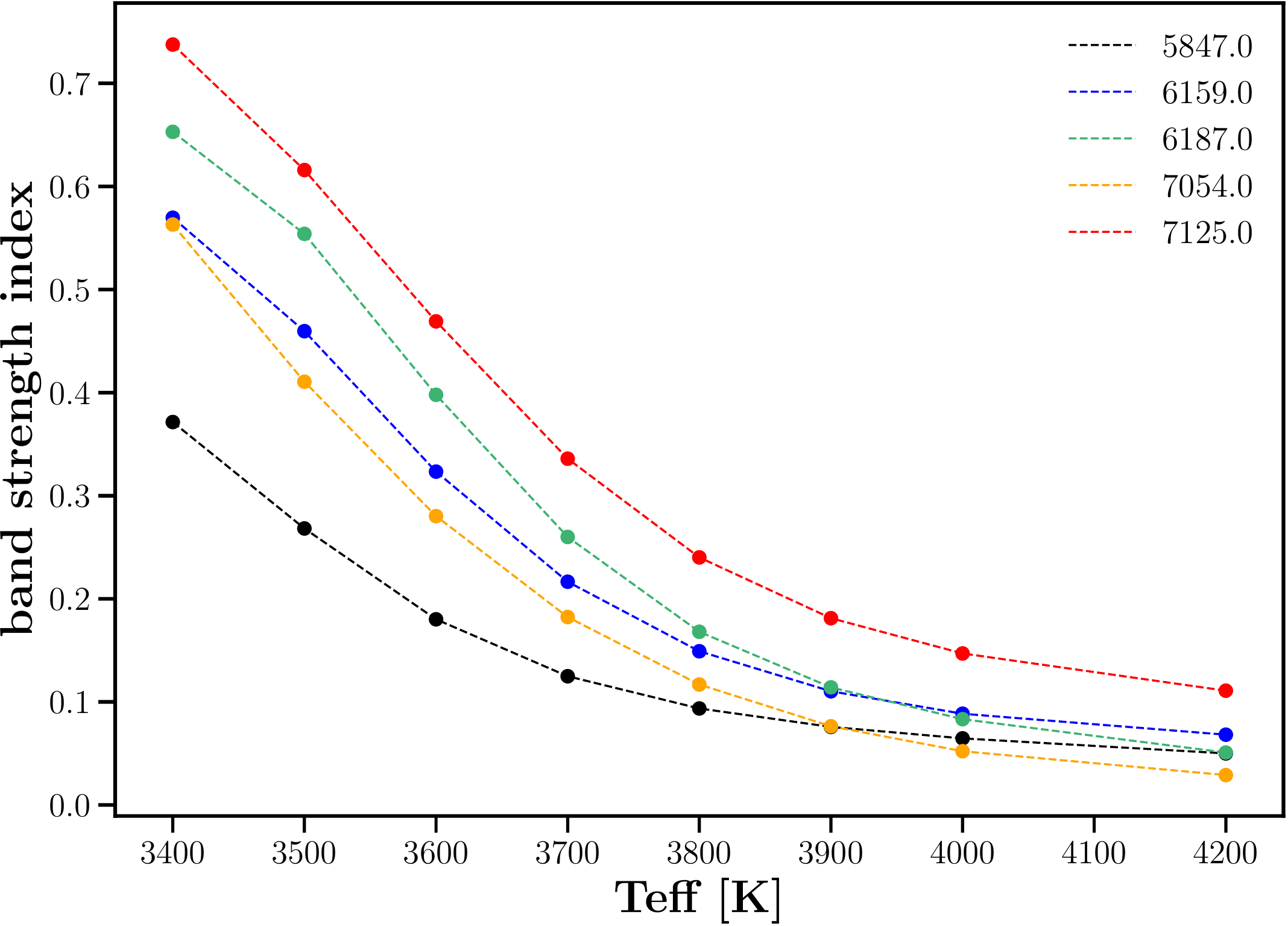}
      \caption{Sensitivity of the depths of TiO bands (i.e., band-strength indices, see Sect.~\ref{Sect:effective_temperatures}) to the effective temperature as computed from the synthetic spectra of 1D MARCS model atmospheres with a fixed $\rm \log g$. The figure clearly shows that the TiO $7125$~\AA\ band is the most sensitive to $T_{\rm eff}$, whereas the TiO $5847$~\AA\ band is the least sensitive.
      }
         \label{Fig:band_sensitivity}
   \end{figure}

   \begin{figure*}
   \centering
   \includegraphics[width=15cm]{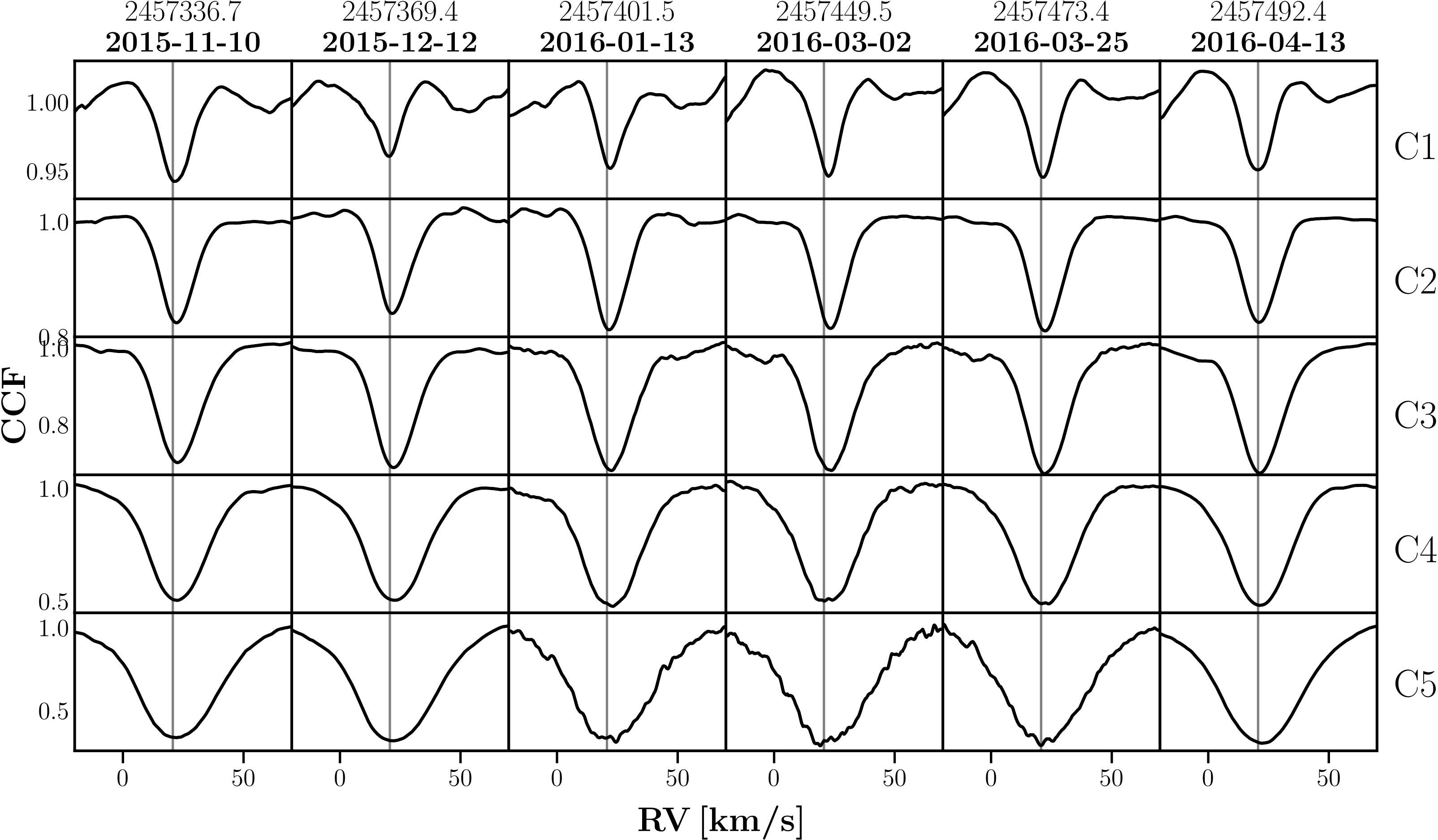}
      \caption{Excerpt of CCF sequence corresponding to time range between JD~2~457~336.7 (November 10, 2015) and JD~2~457~492.4 (April 13, 2016) (compared with the light curve of Betelgeuse in the top panel of Fig.~\ref{Fig:RESULT}). Rows correspond to different masks; mask C1
probes the innermost atmospheric layer, while mask C5 probes the outermost layer. The vertical lines at 20.7~$\rm km \, s^{-1}$ indicate the center-of-mass velocity. 
      }
         \label{Fig:ccf_plot_1}
   \end{figure*}

   \begin{figure*}
   \centering
   \includegraphics[width=15cm]{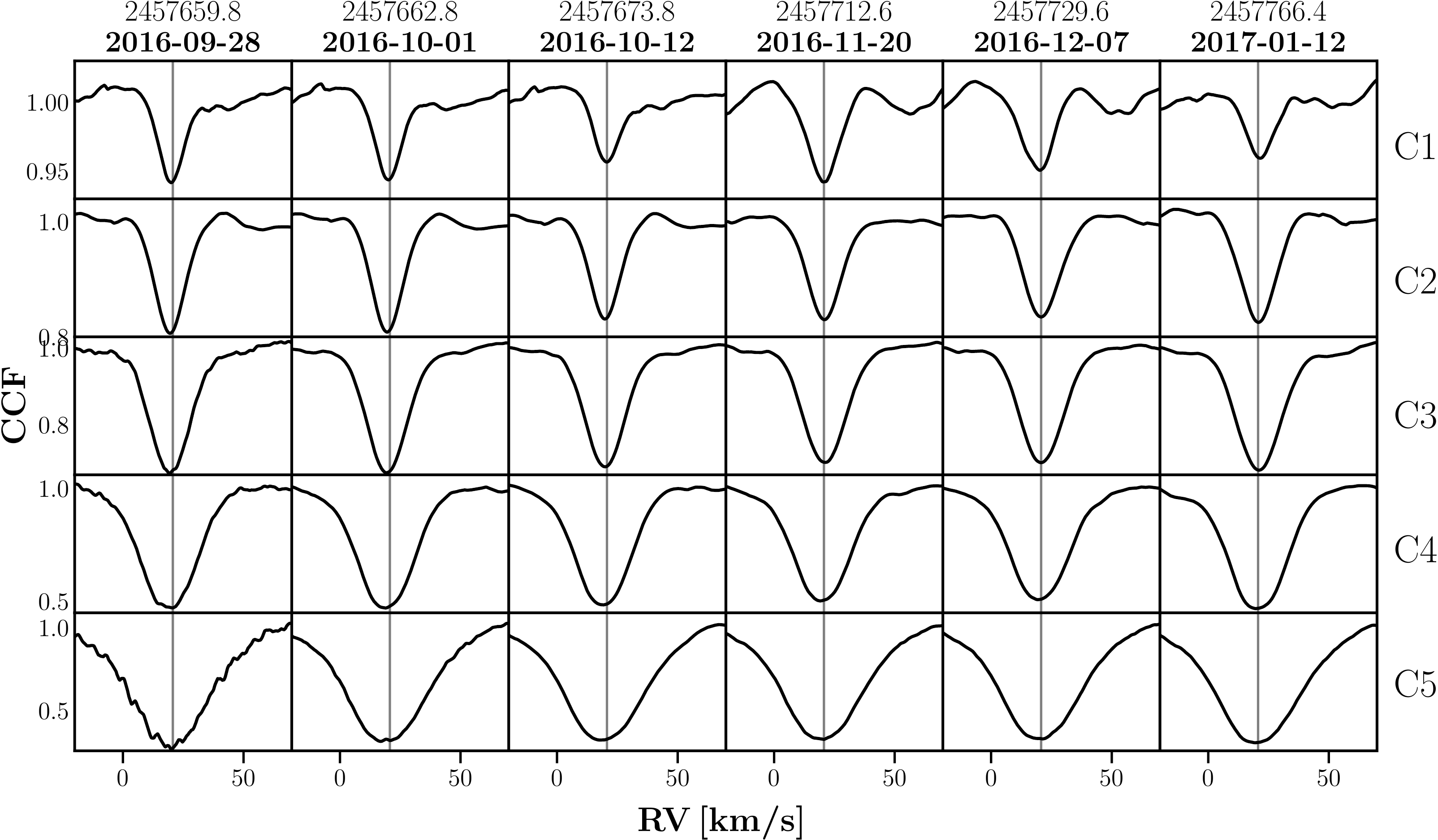}
      \caption{Same as Fig.~\ref{Fig:ccf_plot_1}, but for the time range between JD~2~457~659.8 (September 28, 2016) and JD~2~457~766.4 (January 12, 2017).
      }
         \label{Fig:ccf_plot_2}
   \end{figure*}

   \begin{figure*}
   \centering
   \includegraphics[width=15cm]{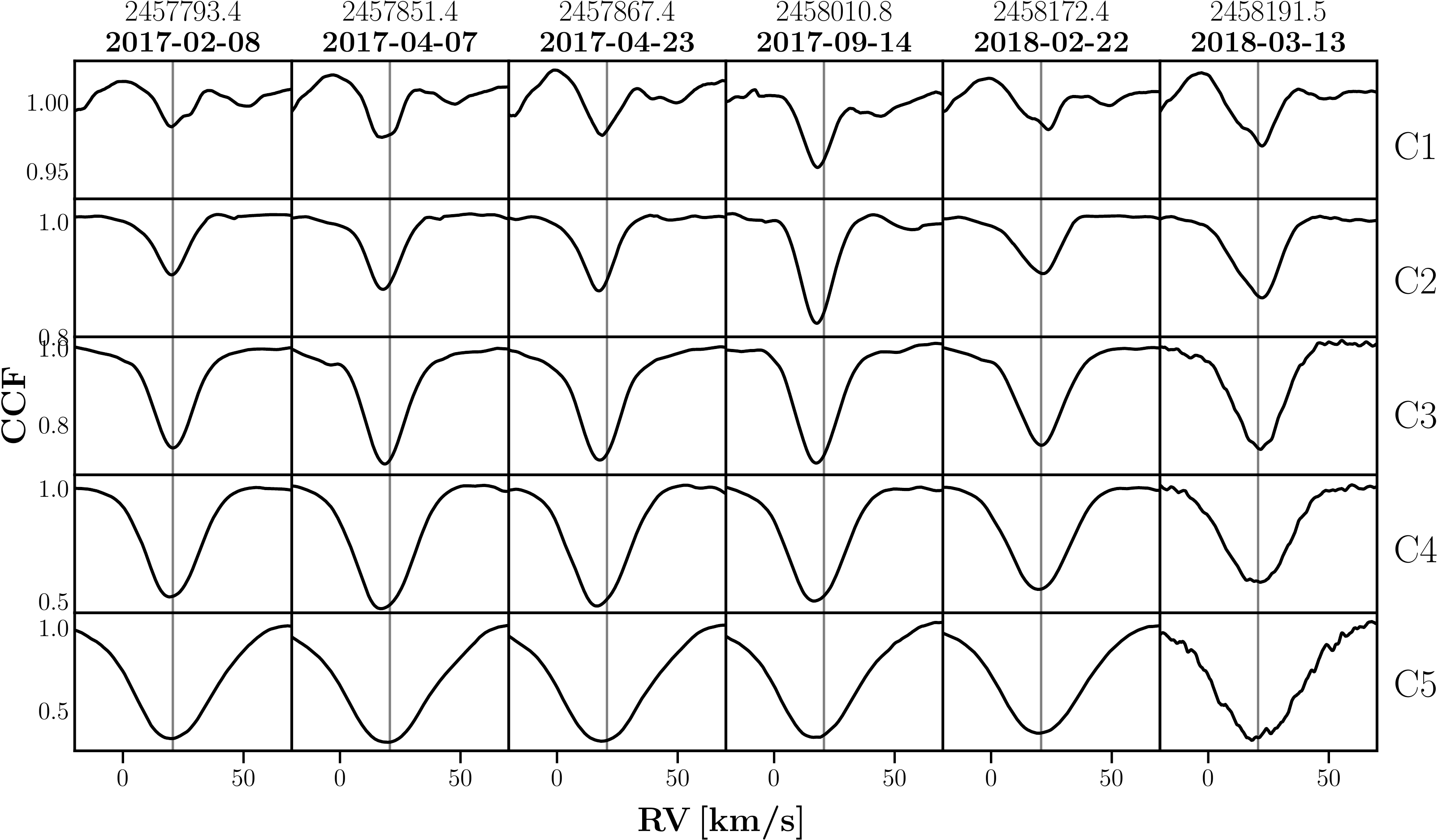}
      \caption{ Same as Fig.~\ref{Fig:ccf_plot_1}, but for time range between JD~2~457~793.4 (February 8, 2017) and JD~2~458~191.5 (March 13, 2018).
      }
         \label{Fig:ccf_plot_3}
   \end{figure*}
   
   \begin{figure*}
   \centering
   \includegraphics[width=15cm]{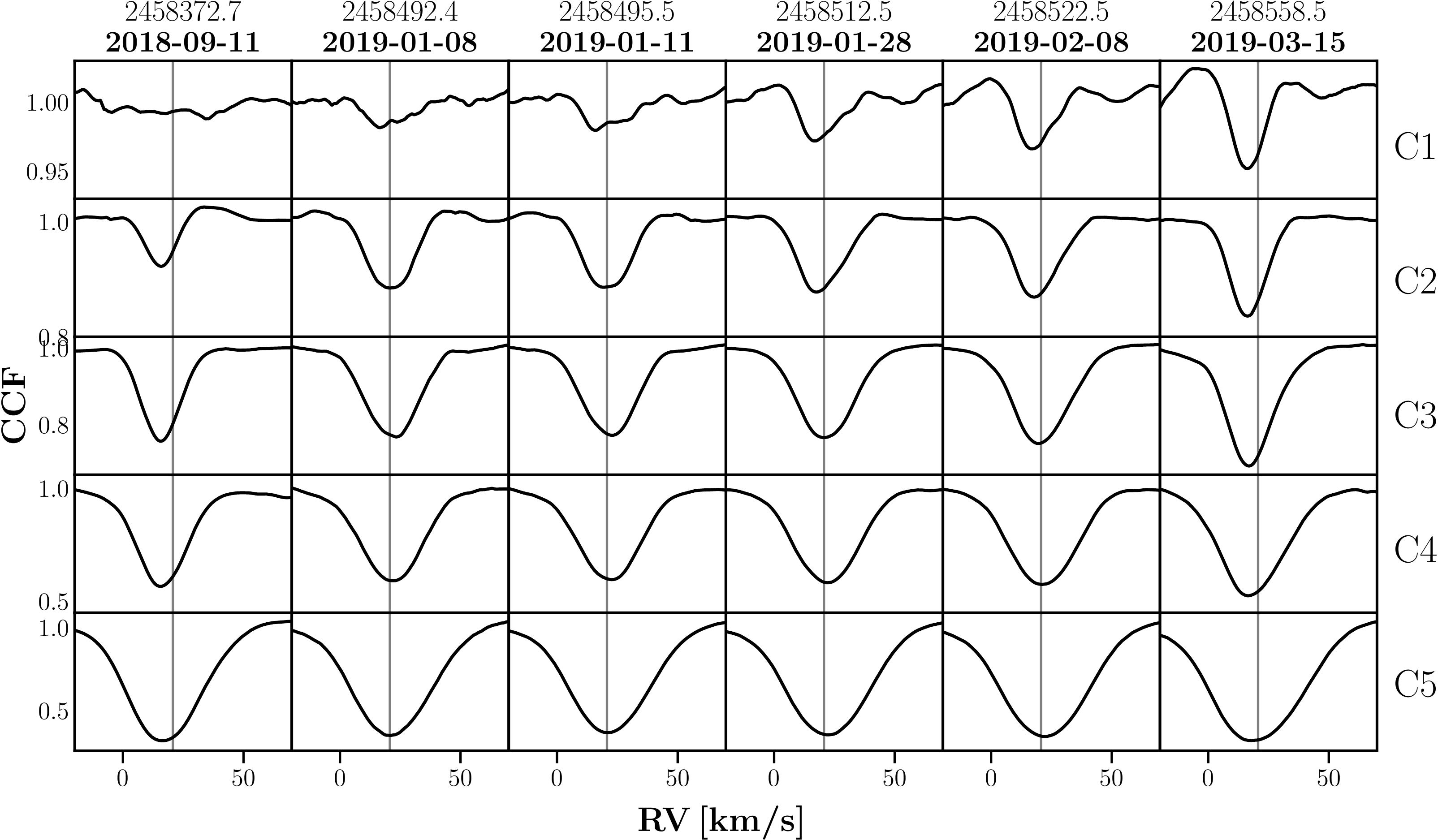}
      \caption{Same as Fig.~\ref{Fig:ccf_plot_1}, but for time range between JD~2~458~372.7 (September 11, 2018) and JD~2~458~558.5 (March 15, 2019).
      }
         \label{Fig:ccf_plot_4}
   \end{figure*}

   \begin{figure*}
   \centering
   \includegraphics[width=15cm]{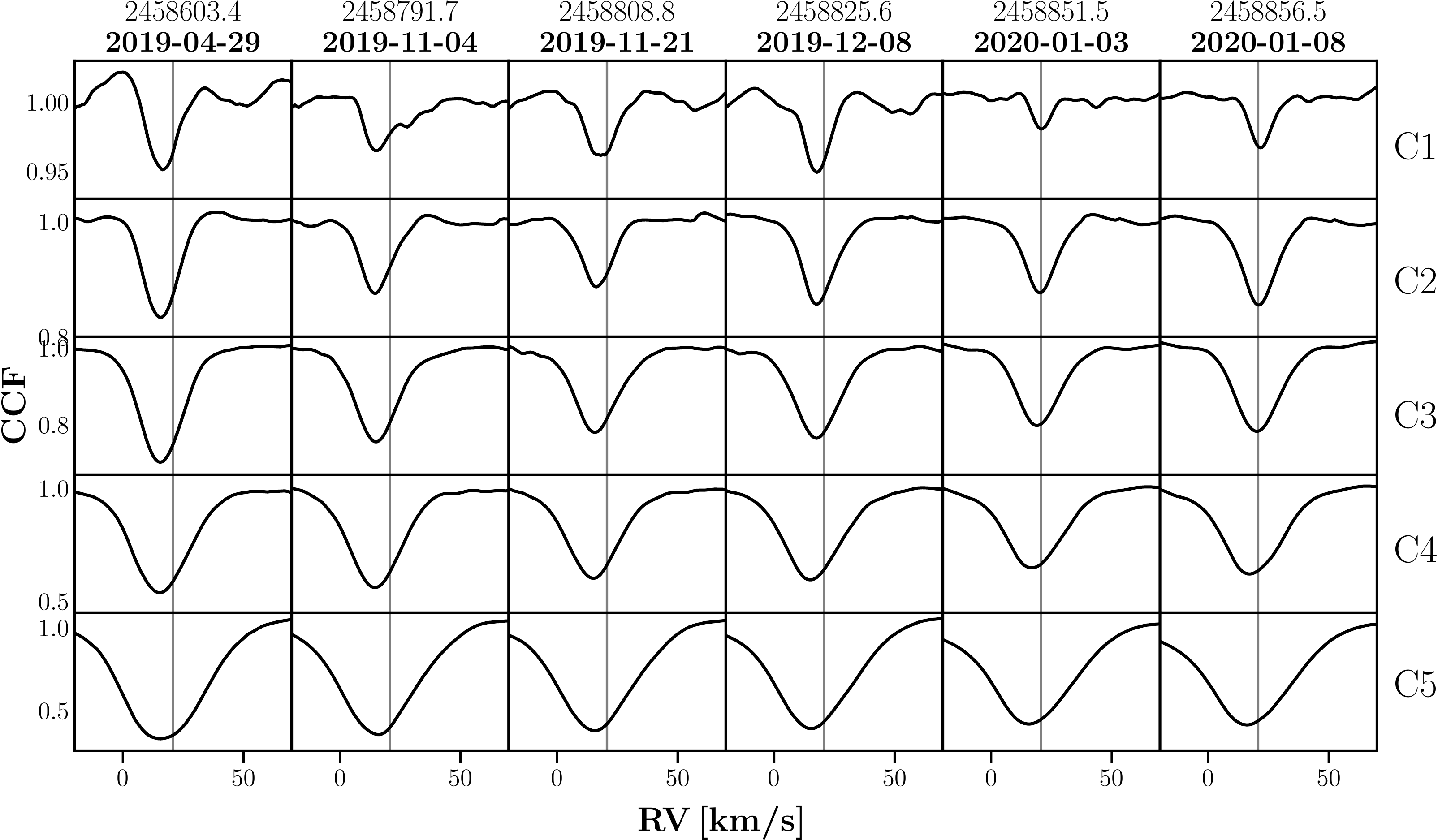}
      \caption{ Same as Fig.~\ref{Fig:ccf_plot_1}, but for time range between JD~2~458~603.4 (April 29, 2019) and JD~2~458~856.5 (January 8, 2020).
      }
         \label{Fig:ccf_plot_5}
   \end{figure*}

   \begin{figure*}
   \centering
   \includegraphics[width=15cm]{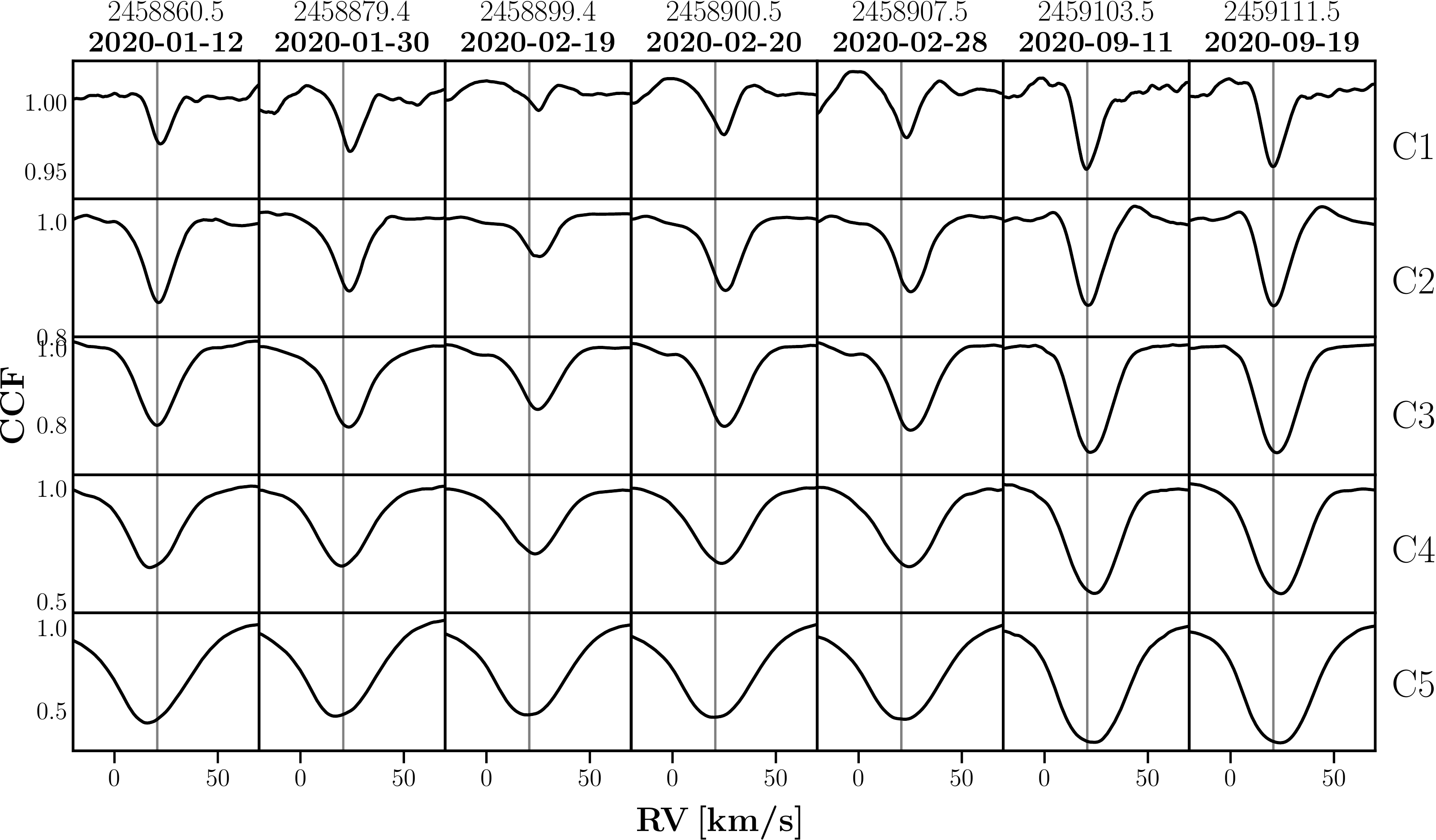}
      \caption{ Same as Fig.~\ref{Fig:ccf_plot_1}, but for time range between JD~2~458~860.5 (January 12, 2020) and JD~2~459~111.5 (September 19, 2020).
      }
         \label{Fig:ccf_plot_6}
   \end{figure*}

\begin{table}[h!]
\begin{center}
\caption[]{TiO temperatures for different observing epochs of Betelgeuse.}
\label{tab:temperatures} 
\begin{tabular}{c c}
\hline \hline
\noalign{\smallskip}
   Date (yy-mm-dd) & TiO temperature [K] \\
   \noalign{\smallskip}
\hline
   \noalign{\smallskip}
2015-11-10 &    3566 \\ 
2015-12-12 &    3585 \\
2016-01-13 &    3599 \\
2016-03-02 &    3602 \\
2016-03-25 &    3633 \\
2016-04-13 &    3647 \\
2016-09-28 &    3638 \\
2016-10-01 &    3633 \\
2016-10-12 &    3626 \\
2016-11-20 &    3620 \\
2016-12-07 &    3623 \\
2017-01-12 &    3620 \\
2017-02-08 &    3618 \\
2017-04-07 &    3646 \\
2017-04-23 &    3644 \\
2017-09-14 &    3599 \\
2018-02-22 &    3606 \\
2019-01-08 &    3586 \\
2019-01-11 &    3586 \\
2019-01-28 &    3586 \\
2019-02-08 &    3595 \\
2019-03-15 &    3643 \\
2019-04-29 &    3610 \\
2019-11-04 &    3577 \\
2019-12-08 &    3564 \\
2020-01-03 &    3562 \\
2020-01-08 &    3561 \\
2020-01-12 &    3561 \\
2020-01-30 &    3553 \\
2020-02-20 &    3542 \\
2020-02-28 &    3539 \\
2020-09-11 &    3545 \\
2020-09-19 &    3544 \\
\noalign{\smallskip}
\hline
\end{tabular}
\end{center}
\end{table}

\end{appendix}

\end{document}